\documentclass{aa5}
\usepackage{graphics}

\begin{document}

\title{ORFEUS echelle spectra: Molecular hydrogen in disk, IVC, and HVC gas 
in front of the LMC}

\author{Hartmut Bluhm \inst{1} \and Klaas. S. de Boer \inst{1} \and Ole Marggraf \inst{1} \and Philipp Richter \inst{1,2}}

\institute{Sternwarte, Universit\"at Bonn, Auf dem H\"ugel 71, 
D-53121 Bonn, Germany
\and 
 Washburn Observatory, University of Wisconsin--Madison, 475 N. Charter Street, Madison, WI 53706, USA}

\date{Received <date> / Accepted <date>}


\offprints{H. Bluhm
\email{hbluhm@astro.uni-bonn.de}
}

\authorrunning{H. Bluhm et al.}
\titlerunning{H$_2$ in disk, IVC, and HVC gas in front of LMC}

\abstract{
 In front of the LMC molecular hydrogen is found in absorption 
near 0 km\,s$^{-1}$, being local disk gas, 
near +60 km\,s$^{-1}$ in an intermediate velocity cloud, 
and near +120 km\,s$^{-1}$, being a high velocity halo cloud. 
 The nature of the gas is discussed based on four {\sc Orfeus} far UV spectra 
of LMC stars 
 and including data 
from the ground and from the IUE satellite.
 The local gas is cool and, given a span of sight lines of only 2.5$^\circ$, 
rather fluffy.
 The  fractional abundance of H$_2$ varies from $\log f=\log [N(\rm{H}_2)/(2\cdot N(\rm{H}_2)+N(\rm{\ion{H}{i}}))]=-5.4$ to $-3.3$.
 Metal depletions (up to $-1.7$~dex for Fe) are typical for galactic disk gas. 
 In the intermediate and high velocity gas an apparent underabundance of neutral oxygen points to an ionization level of the gas of about $90$~\%.
 H$_2$ is detected in intermediate and high velocity gas towards \object{HD~269546}.
 In the intermediate velocity gas we find an H$_2$ column density of $\log(N)\simeq15.6$. 
 The H$_2$ excitation indicates that the line of sight samples a cloud at a temperature below $150$~K.
 Column densities are too small to detect the higher UV pumped excitation levels.  
 The high velocity H$_2$ ($\log(N)\simeq15.6$) is highly excited and probably exposed to a strong radiation field. 
 Its excitation temperature exceeds $1000$~K.
 Due to the radial velocity difference between the halo gas and the Milky Way disk, the unattenuated disk radiation is available for H$_2$ excitation in the halo. 
 We do not find evidence for an intergalactic origin of this gas; a galactic as well as a Magellanic Cloud origin is possible.  
}

\maketitle

\section{Gas in front of the LMC}

Shortly after the launch of the International Ultraviolet Explorer 
({\sc iue} ) satellite, 
gas absorbing at small and large radial velocity was found in spectra of 
stars in the Magellanic Clouds (Savage \& de Boer \cite{savdB}). 
The absorption by Milky Way disk gas, of course near 0 km\,s$^{-1}$, 
included species like C\,{\sc i}, O\,{\sc i}, Si\,{\sc ii}, Fe\,{\sc ii}, 
Mg\,{\sc ii} and the like. 
Gas in the Large Magellanic Cloud (LMC) and Small Magellanic Cloud (SMC) 
is present in absorption at the velocity of each of these galaxies. 

In the direction of the LMC gas was found also at intermediate velocities, 
near +60 km\,s$^{-1}$ and near +120~km\,s$^{-1}$ (Savage \& de Boer \cite{savdB}). 
Both velocities were seen in absorption by the normal neutral species. 
Absorption by highly ionized species such as Si\,{\sc iv} and C\,{\sc iv} 
was present at non zero velocities as well. 
Clearly, an intermediate velocity cloud (IVC) and 
a high velocity cloud (HVC) had been detected. 

Since then numerous further observations in the visual and the ultraviolet 
of other Magellanic Cloud stars confirmed these interstellar absorptions. 
In addition, probing at 21 cm outside the face of the LMC showed that 
in particular the gas at +120 km\,s$^{-1}$ has a rather wide angular spread
(de Boer et\,al.\, \cite{dBmorras}). 
The latter observations gave strong support for the claim that 
the +120 km\,s$^{-1}$ gas is part of the galactic halo. 

A new wavelength domain was opened for the observation of stars 
in the Magellanic Clouds with the {\sc orfeus} space shuttle mission. 
The echelle spectrograph on board {\sc orfeus} produces spectra between 
900 and 1400 \AA\ with a spectral resolution of $\simeq$30 km\,s$^{-1}$ 
(see Barnstedt et\,al.\, \cite{barnstedt}). 
Thus interstellar absorption lines of molecular hydrogen H$_2$ 
(and also of O\,{\sc vi}) became accessible.

 Analyses of the LMC part of the {\sc orfeus} absorption spectra have 
resulted in the detection of H$_2$ (de Boer et\,al.\, \cite{dBrichter}) 
and in a study of the H$_2$/CO ratio (Richter et\,al.\, \cite{richter99b}). 
 A full analysis of {\sc orfeus} H$_2$ observations in both Magellanic Clouds is presented 
by Richter (\cite{richter2000}).
 
 Here we will analyse the absorption by H$_2$ and metals 
in the neutral gas in the foreground of the LMC using {\sc orfeus} spectra (see Table \ref{targets}). 
 The data were obtained during the mission of Nov./Dec. 1996.

 Detailed inspection of the {\sc orfeus} LMC star spectra showed the existence of absorption by H$_2$ at intermediate velocities, towards \object{HD~269546}.
 This is the third IVC seen in H$_2$. 
 The first IVC with H$_2$ was found towards HD~93521 by 
Gringel et al. (\cite{gringel}), the second towards PG~0804+761 by Richter et al. (\cite{richter2000b}).
 The discovery of H$_2$ in the +120~km\,s$^{-1}$ HVC was 
presented by Richter et\,al.\, (\cite{richter99a}). 

%
\section{Data and data reduction}

The data were preanalysed by the {\sc orfeus} team in T\"ubingen and 
provided in the form of spectra order by order (Barnstedt et\,al.\, \cite{barnstedt}).  
The interstellar absorption lines exhibit in part terrible blending, 
due to the rather densely packed absorptions by H$_2$ 
together with the large ($\simeq 300$ km\,s$^{-1}$) range in velocity 
covered by each absorption.  
Given the considerable blending, 
not each velocity component could be identified for all transitions 
individually.
Only blend free lines and blend free portions of absorption profiles 
have been selected for the present study. 
The tabular material presented may therefore give the impression 
of inhomogeneity.

\begin{figure*}
\resizebox{\hsize}{!}{\includegraphics{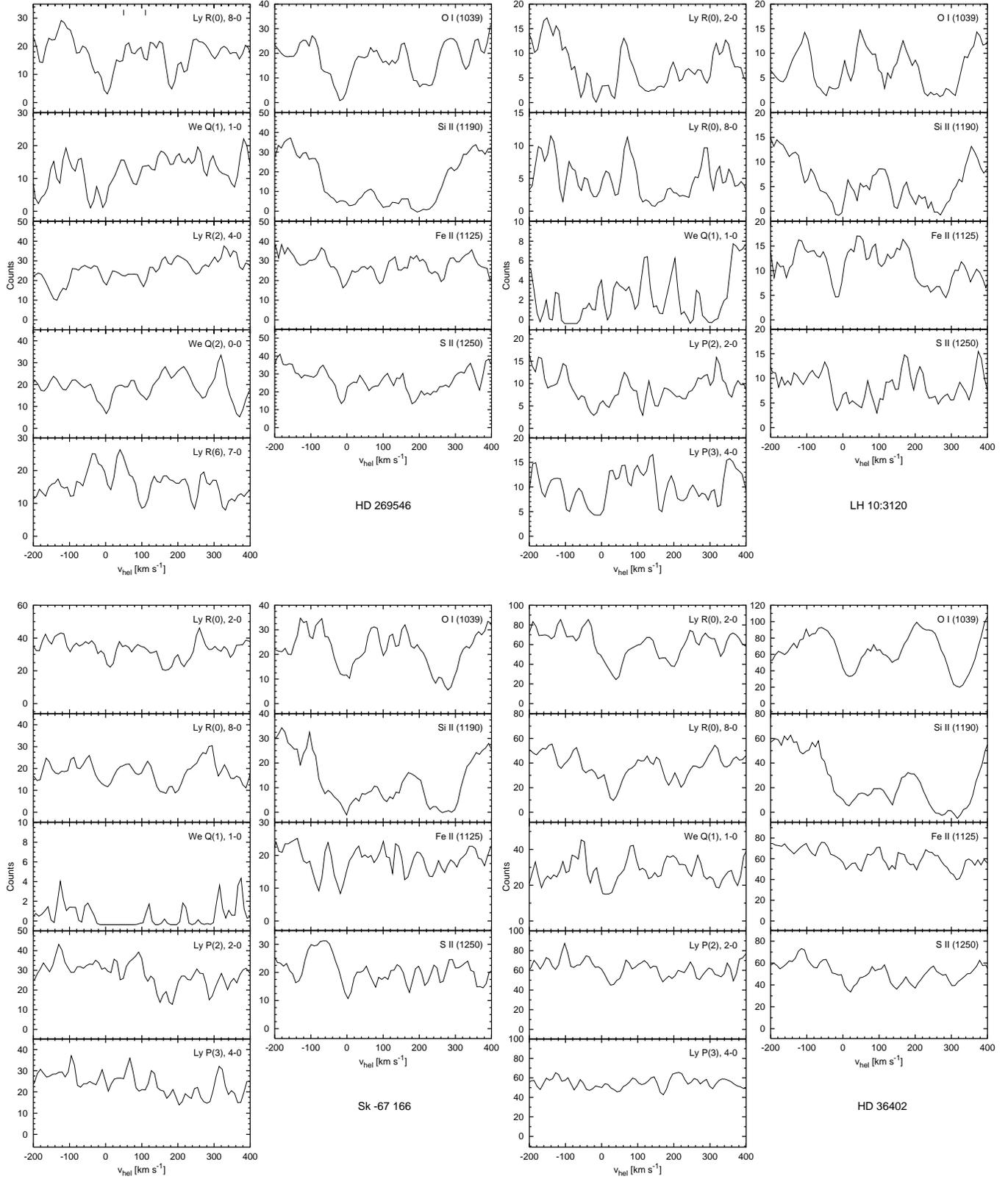}}
\hfill
\caption[]{For each line of sight a sample of five H$_2$ and four metal lines as measured with {\sc Orfeus} is shown. Additional transitions within the range of the plots are: Ly R(1),8-0, Ly P(4),9-0 [Ly R(0),8-0]; Ly P(5),11-0, Ly R(3),10-0), We R(3),1-0, We R(0),1-0, We R(1),1-0 [We Q(1),1-0]; Ly P(1),4-0 [Ly R(2),4-0]; We R(4),0-0, We R(3),0-0 [We Q(2),0-0]; Ly P(3),6-0, \ion{Cl}{i} [Ly R(6),7-0]; \ion{Fe}{ii}, Ly R(3),2-0 [Ly P(2),2-0]; Ly R(2),5-0 [\ion{O}{i}]; \ion{S}{iii} [\ion{Si}{ii} 1190~{\AA}]; \ion{P}{ii} [\ion{Fe}{ii} 1125~{\AA}]     
}
\label{spectra}
\end{figure*}

 The target data are listed in Table \ref{targets}.
 Examples of absorption profiles, both for H$_2$ as well as metal lines, 
are given in Fig.\,\ref{spectra} for each of the stars. 

 Spectra had also been obtained with the {\sc iue} satellite, except for \object{LH~10:3120}. 
 We have included these spectra (from the public archive) in our analysis. 
 Note that the {\sc iue} spectrum SWP~13347 is not of \object{HD~269546} (here the {\sc iue} archive is at fault; see Chu et al. \cite{chu}). 
 A preliminary analysis of the velocity and Si\,{\sc ii} column density 
in the IVC and HVC in front of the LMC 
based on {\sc iue} spectra of many LMC stars 
has been presented by Wierig \& de~Boer (\cite{wierig}). 

The intercomparison of {\sc orfeus} and {\sc iue} data 
in the spectral range common to both instruments 
shows that the absorption profiles agree well. 
We have therefore combined the data from {\sc orfeus} with those 
available from the {\sc iue} and can so obtain a more complete picture 
for the nature of the gas in the three velocity components studied. 

Data of H\,{\sc i} emission at 21 cm are available from the literature 
(McGee et\,al.\, \cite{mcgee83}, McGee \& Newton \cite{mcgee86}, de~Boer et\,al.\, \cite{dBmorras}).

 At this point we have to dwell briefly on the velocity scale 
used for the data. 
 The zero point of the velocity scale of the {\sc orfeus} spectra  
is heliocentric 
for stars whose position in the sky was perfectly known and 
which were observed at the centre of the {\sc orfeus} aperture 
(Barnstedt et al. \cite{barnstedt}). 
 Neither of the requirements must have been fulfilled for our stars. 
 A similar situtation holds for the {\sc iue} data. 
 In the four analysed {\sc orfeus} spectra the zero points of the velocity scales should be correct within 10~km\,s$^{-1}$.

\begin{table}
\caption[]{LMC stars observed with the {\sc orfeus} echelle spectrograph}
\label{targets}
\setlength{\tabcolsep}{0.6mm}
\begin{flushleft}
\begin{tabular}{lcclcc}
\hline\noalign{\smallskip}
 Star & $l$\ \ \ \ \ \ \ \ $b$ & $V$ & Sp.T. & $E(B$-$V)$ & exp.time\\
     & [degree] & [mag] & &  & [sec] \\
\noalign{\smallskip}\hline\noalign{\smallskip}
\object{LH~10:3120} & 277.2 $-$36.1 & 12.80 & O5.5V & \hspace{0.25cm}$0.17$ & 6528\\
\object{Sk~-67\,166} & 277.8 $-$32.5 & 12.27 & O5e & \hspace{0.25cm}$0.09$ & 6188\\
\object{HD~36402} & 277.8 $-$33.0 & 11.50 & OB+WC5 & $\leq$$0.02$ & 6764\\
\object{\object{HD~269546}} & 279.3 $-$32.8 & 11.30 & B3+WN3 & $\leq$$0.02$ & 6388\\
\noalign{\smallskip}\hline\noalign{\smallskip}
\end{tabular}

References for $E(B-V)$ values: \object{LH~10:3120}: Parker et al. (\cite{parker92}), \object{Sk~-67\,166}: Wilcots et al. (\cite{wilcots96}), \object{HD~36402}: de Boer \& Nash (\cite{dBnash}), \object{HD~269546}: Vacca \& Torres-Dodgen (\cite{vaccatorres90}).
 We use the names by which the LMC stars are best known. Other names are:
 \object{Sk~-67\,166} = \object{HD~269698}, \object{HD~36402} = \object{Sk~-67\,104}, \object{HD~269546} = \object{Sk~-68\,82}
\end{flushleft}
\end{table}

%
\section{Toward column densities and gas properties}

\begin{table}
\caption[]{Analysed metal absorption lines. Measurements in {{\sc orfeus}} and {{\sc iue}} spectra are denoted by {\sl o} and {\sl i}, respectively }
\label{lines}
\begin{tabular}{lp{1.5cm}p{1.5cm}p{1.5cm}l}
\hline\noalign{\smallskip}
Ion & \multicolumn{4}{l}{Transition (wavelength in \AA)}\\  
\noalign{\smallskip}\hline\noalign{\smallskip}
\ion{Si}{ii} & 1190 {\sl i,o}& 1193 {\sl i,o}& 1260 {\sl i,o}& 1304 {\sl i,o}\\
             & 1526 {\sl i} & 1808 {\sl i} &      &      \\
\ion{Fe}{ii} & 1125 {\sl o} & 1142 {\sl o} & 1144 {\sl o}& 2344 {\sl i} \\
             & 2374 {\sl i} & 2382 {\sl i} & 2586 {\sl i}& 2600 {\sl i} \\
\ion{S}{ii}  & 1250 {\sl i,o} & 1253 {\sl i,o} & 1259 {\sl i,o} &      \\
\ion{O}{i}   & 936  {\sl o} & 976 {\sl o} & 1039 {\sl o} & 1302 {\sl i,o} \\
\ion{Mg}{i}  & 2852 {\sl i} &      &      &      \\
\ion{Mg}{ii} & 2796 {\sl i} & 2803 {\sl i} &      &      \\         
\noalign{\smallskip}\hline\noalign{\smallskip}
\end{tabular}
\end{table}

Each velocity component was analysed in the following manner:
 
 The equivalent widths were measured by trapezium or Gaussian fits.
 Their uncertainties are calculated from the estimated error in the choice of the continuum and the photon statistics, which were taken into consideration using a formula by Jenkins et al. (\cite{jenkins}).
  
 For H$_2$ we constructed (using $f$-values from Morton \& Dinerstein \cite{mortdiner}) 
the curve of growth for each $J$-level independently. 
 These curves generally have the same $b$-value. 
 Column densities then follow, also for those levels for which 
possibly only one line could be measured. 

 The various rotational levels of H$_2$ can be excited through collisions 
with other particles, leading mostly to population of the lower $J$ levels, 
and through photons, leading mostly to population of the higher $J$ levels 
(Spitzer \& Zweibel \cite{spitzweib}). 
 The latter process is also called UV pumping. 
 We have fitted a Boltzmann distribution to the column densities derived for the rotational states.
 This results, for the low $J$ levels, 
in a value for the kinetic temperature of the gas. 
 For the high $J$ levels the fit produces a value merely indicating 
the level of UV pumping. 
 The nature of the high level excitation can, in principle, 
be used to get information about the FUV radiation density 
(Wright \& Morton \cite{wrightmort}). 

 For each atomic species ($f$-values from the compilation by Morton \cite{morton} and from the review by Savage \& Sembach \cite{savsem}) 
a curve of growth was constructed, too. 
 The total $b$-value is mostly based on the curve of growth 
of Si\,{\sc ii} and Fe\,{\sc ii}. 
 Column densities were derived in the usual manner. 
 For those species having too few absorption lines the column density 
was derived by shifting the data to the adopted curve of growth.

 The given errors of the column densities are based on the errors of the equivalent widths and the estimated uncertainty of the $b$-value.  

 For \object{LH~10:3120} no {{\sc iue}} spectrum is available and the {{\sc orfeus}} spectrum has a relatively low countrate (mostly between 10 and 20 counts).
 For that reason only a few strong lines could be measured.

%
\section{Densities and temperature in the clouds}

 Above we have given a general description of the methods of analysis 
employed to derive velocities, column densities, temperatures, 
excitation conditions, etc. from the data. 
 Here we will discuss the results for each velocity component 
as derived from the available lines of sight. 
 H$_2$ equivalent widths and column densities are listed in Tables \ref{H2eqw_disk}, \ref{H2eqw_halo}, \ref{H2CD_local}, and \ref{H2CD_halo}.   
 Metal and \ion{H}{i} column densities are given in Table \ref{colden}.

\begin{table*}
\caption[]{H$_2$ equivalent widths of galactic disk gas in the spectra of LMC stars}
\label{H2eqw_disk}
\begin{flushleft}
\begin{tabular}{lrlllll}
\hline\noalign{\smallskip}
 & \multicolumn{1}{c}{$\lambda$~[\AA]} & \multicolumn{5}{c}{$W_\lambda$~[m\AA]} \\
\noalign{\smallskip}\cline{3-7}\noalign{\smallskip}
 &                                     & \object{\object{HD~269546}} & LH 10:3120 & \object{Sk~-67\,166} & \object{HD~36402} (A) & \object{HD~36402} (B) \\
\noalign{\smallskip}\hline\noalign{\smallskip}
Ly R(0), 0-0 & 1108.128 & n          & $48\pm24$  & $<15$     & n         & $49\pm14$ \\
Ly R(0), 2-0 & 1077.138 & $121\pm19$ & $114\pm24$ & $28\pm8$  & $26\pm13$ & $87\pm17$ \\
Ly R(0), 4-0 & 1049.366 & b?,n       & b?,n       & b?,n      & $28\pm8$  & b?,n      \\
Ly R(0), 8-0 & 1001.826 & $104\pm19$ & $94\pm33$  & n         & $26\pm12$ & $102\pm17$ \\
Ly R(0), 10-0 & 981.441 & n,b?      & $143\pm28$ & n         & n         & n         \\
Ly R(0), 12-0 & 962.978 & b?        & f          & $31\pm15$ & $23\pm12$ & b?,n      \\
\noalign{\smallskip}\hline\noalign{\smallskip}
Ly R(1), 0-0 & 1108.634 & $58\pm18$  & n          & $<16$     & $<13$     & $32\pm14$ \\
We Q(1), 0-0 & 1009.772 & $96\pm18$  & n          & $34\pm14$ & $25\pm7$  & $51\pm12$ \\
Ly R(1), 1-0 & 1092.732 & b (LR0,1-0,HVC) & n?    & $15\pm9$  & n         & --        \\
We Q(1), 1-0 & 986.798  & $87\pm20$  & $49\pm29$  & f         & $34\pm8$  & $50\pm9$  \\
Ly R(1), 2-0 & 1077.698 & $60\pm18$  & n          & $25\pm9$  & $19\pm9$  & --        \\
We Q(1), 3-0 & 947.425  & b?         & f          & b?        & $33\pm14$ & b?        \\
Ly R(1), 4-0 & 1049.958 & $97\pm18$  & f          & $54\pm11$ & $21\pm10$ & $72\pm12$ \\
Ly R(1), 7-0 & 1013.434 & $85\pm24$  & $98\pm29$  & n         & n         & n         \\
Ly R(1), 8-0 & 1002.457 & n          & $75\pm29$  & $40\pm11$ & $35\pm13$ & $56\pm14$ \\
Ly R(1), 12-0 & 963.609 & f         & $82\pm23$  & --        & n         & f         \\
Ly R(1), 13-0 & 955.067 & f         & b (LR0)    & $<32$     & $31\pm12$ & $65\pm18$ \\
\noalign{\smallskip}\hline\noalign{\smallskip}
Ly P(2), 0-0 & 1112.495 & b (FeII, CI) & b        & $<32$     & b         & n         \\
Ly P(2), 1-0 & 1096.439 & n          & $69\pm25$  & n         & n         & $<15$     \\
Ly P(2), 2-0 & 1081.265 & $40\pm14$  & $76\pm24$  & $14\pm11$ & $<11$     & $18\pm9$  \\
Ly P(2), 6-0 & 1028.103 & $47\pm18$  & f          & $36\pm22$ & f         & f         \\
Ly P(2), 7-0 & 1016.472 & $61\pm16$  & $42\pm22$  & n         & $17\pm8$  & $16\pm9$  \\
Ly P(2), 8-0 & 1005.397 & $33\pm19$  & $56\pm23$  & n         & n         & b         \\
\noalign{\smallskip}\hline\noalign{\smallskip}
Ly P(3), 1-0 & 1099.788 & $26\pm12$  & --         & $17\pm12$ & n         & $<32$     \\
Ly P(3), 2-0 &  1084.559 & b          & $39\pm28$  & n         & n         & b         \\
We P(3), 2-0 & 970.560  & n,f        & $44\pm28$  & n         & f         & f         \\
Ly P(3), 4-0 & 1056.469 & $50\pm16$  & --         & $23\pm14$ & $<22$     & $13\pm6$  \\
Ly P(3), 5-0 & 1043.498 & b          & b          & b         & b         & $20\pm8$  \\
Ly P(3), 15-0 & 944.331 & n         & $41\pm24$  & $23\pm21$ & $<20$     & n         \\
\noalign{\smallskip}\hline\noalign{\smallskip}
Ly P(4), 0-0 & 1120.247 & $<20$      & n          & --        & --        & $12\pm7$  \\
Ly P(4), 2-0 & 1088.794 & $27\pm13$  & $34\pm20$  & --        & --        & n         \\
Ly P(4), 3-0 & 1074.313 & n          & $22\pm17$  & --        & --        & n         \\
Ly P(4), 5-0 & 1047.554 & $44\pm15$  & n          & $<18$     & --        & n         \\
\noalign{\smallskip}\hline\noalign{\smallskip}
Ly P(5), 1-0 & 1109.313 & n          & --         & $<21$     & --        & $<16$     \\
We P(5), 1-0 & 997.640  & n          & $<26$      & --        & --        & --        \\
Ly P(5), 8-0 & 1017.009 & $35\pm14$  & --         & --        & --        & --        \\
Ly P(5), 14-0 & 960.265 & $<26$     & --         & --        & --        & --        \\
\noalign{\smallskip}\hline\noalign{\smallskip}
\end{tabular}

Wavelengths are taken from Morton \& Dinerstein (\cite{mortdiner}).
Some equivalent widths have not been measured due to: (n) noise, (f) low flux, (b) blend. Some blends could not be identified
\end{flushleft}
\end{table*}

\begin{table*}
\caption[]{Halo H$_2$ equivalent widths towards the LMC stars}
\label{H2eqw_halo}
\begin{flushleft}
\begin{tabular}{lrlllllllll}
\hline\noalign{\smallskip}
 & \multicolumn{1}{c}{$\lambda$~[\AA]} & \multicolumn{8}{c}{$W_\lambda$~[m\AA]} \\
\noalign{\smallskip}\cline{3-10}\noalign{\smallskip}
 & & \multicolumn{2}{l}{\object{\object{HD~269546}}} & \multicolumn{2}{l}{LH 10:3120} & \multicolumn{2}{l}{\object{Sk~-67\,166}} & \multicolumn{2}{l}{\object{HD~36402}} \\
 & & IVC & HVC              & IVC & HVC       & IVC & HVC       & IVC & HVC \\ 
\noalign{\smallskip}\hline\noalign{\smallskip}
Ly R(0), 1-0 & 1092.194 & $59\pm28$ & b (LR1,1-0) & $<16$ & --    & --    & --    & --    & --    \\
Ly R(0), 2-0 & 1077.138 & $76\pm22$ & b (LR1,2-0) & --    & --    & $<16$ & --    & --    & --    \\
Ly R(0), 4-0 & 1049.366 & n,b?      & $<30$       & --    & $<60$ & --    & $<14$ & --    & $<7$  \\
Ly R(0), 8-0 & 1001.826 & $79\pm17$ & b? (LR1)    & --    & --    & $<25$ & $<19$ & $<12$ & $<12$ \\
Ly R(0), 10-0 & 981.441 & $77\pm23$ & b? (LR1)   & $<17$ & $<56$ & --    & --    & $<18$ & $<16$ \\
\noalign{\smallskip}\hline\noalign{\smallskip}
Ly R(1), 1-0 & 1092.732 & $46\pm25$ & b (LR0,1-0) & $<26$ & --    & --    & --    & --    & --    \\
We Q(1), 1-0 & 986.798  & b (LR3,10-0,LMC) & $75\pm22$ & -- & --  & --    & --    & --    & $<26$ \\
Ly R(1), 2-0 & 1077.698 & $63\pm19$ & b (LR0,2-0) & $<61$ & --    & $<26$ & --    & $<13$ & --    \\
Ly P(1), 5-0 & 1038.156 & $54\pm31$ & b (LR0)     & --    & --    & $<17$ & --    & --    & --    \\
Ly R(1), 8-0 & 1002.457 & $71\pm22$ & $34\pm19$   & $<63$ & $<47$ & $<21$ & $<29$ & $<11$ & --    \\
\noalign{\smallskip}\hline\noalign{\smallskip}
We Q(2), 0-0 & 1010.941 & $<38$     & $65\pm22$   & --    & $<36$ & $<12$ & $<12$ & $<9$  & $<9$  \\
Ly R(2), 4-0 & 1051.497 & n         & $54\pm23$   & $<30$ & --    & $<22$ & --    & $<7$  & $<8$  \\
Ly P(2), 7-0 & 1016.472 & n         & $<57$       & $<25$ & $<35$ & $<21$ & $<10$ & $<13$ & $<15$ \\
Ly P(2), 11-0 & 975.343 & n        & $<42$       & --    & --    & --    & --    & --    & --    \\
\noalign{\smallskip}\hline\noalign{\smallskip}
Ly P(3), 3-0 & 1070.142 & $<36$     & $33\pm17$   & $<40$ & --    & --    & $<23$ & $<16$ & --    \\
Ly R(3), 4-0 & 1053.976 & b?        & $47\pm23$   & $<75$ & --    & $<14$ & $<20$ & --    & $<19$ \\
Ly P(3), 4-0 & 1056.469 & n         & n           & $<18$ & $<18$ & --    & --    & --    & --    \\
Ly P(3), 5-0 & 1043.498 & $39\pm27$ & n           & --    & $<33$ & --    & --    & $<18$ & --    \\
Ly R(3), 8-0 & 1006.418 & n,b?      & $63\pm30$   & --    & $<63$ & $<20$ & --    & --    & $<28$ \\
\noalign{\smallskip}\hline\noalign{\smallskip}
Ly R(4), 4-0 & 1057.379 & --        & $46\pm23$   & --    & --    & --    & --    & --    & --    \\
\noalign{\smallskip}\hline\noalign{\smallskip}
Ly P(5), 4-0 & 1065.594 & --        & $<31$       & --    & --    & --    & --    & --    & --    \\
\noalign{\smallskip}\hline\noalign{\smallskip}
Ly P(6), 3-0 & 1085.382 & --        & $<54$       & --    & --    & --    & --    & --    & --    \\
Ly R(6), 7-0 & 1030.064 & --        & $65\pm25$   & --    & --    & --    & --    & --    & --    \\
\noalign{\smallskip}\hline\noalign{\smallskip}
\end{tabular}

Wavelengths are taken from Morton \& Dinerstein (\cite{mortdiner}).
Some equivalent widths have not been measured due to: (n) noise, (f) low flux, (b) blend. Some blends could not be identified
\end{flushleft}
\end{table*}

\begin{table*}
\caption[]{Local gas H$_2$ column densities $\log N$ [cm$^{-2}$] and 
          H$_2$ excitation temperatures towards the LMC stars. $v_{\rm hel}$ is the heliocentric velocity from the {\sc orfeus} data, $b$ is the Doppler parameter of the curves of growth}
\label{H2CD_local}
\begin{flushleft}
\begin{tabular}{llllll}
\hline\noalign{\smallskip}
J & \object{\object{HD~269546}} & LH 10:3120 & \object{Sk~-67\,166} & \object{HD~36402} (A) & \object{HD~36402} (B)\\
\noalign{\smallskip}\hline\noalign{\smallskip}
$v_{\mathrm{hel}}$ [km\,s$^{-1}$] & $+5\pm4$ & $-16\pm6$ & $+7\pm5$ & $-4\pm5$ & $+35\pm4$ \\
\noalign{\smallskip}\hline\noalign{\smallskip}
$b$~[km\,s$^{-1}$] & 3-4 & 4-6 & 4-6 & 2-4 & 4-6 \\
\noalign{\smallskip}\hline\noalign{\smallskip}
0 & $17.10^{+0.15}_{-0.25}$ & $17.00^{+0.50}_{-0.90}$ & $14.55^{+0.25}_{-0.30}$ & $14.70^{+0.90}_{-0.55}$ & $15.85^{+0.35}_{-0.45}$\\
1 & $16.70^{+0.35}_{-0.40}$ & $16.20^{+0.60}_{-0.60}$ & $14.60^{+0.25}_{-0.25}$ & $14.85^{+0.35}_{-0.30}$  & $15.45^{+0.30}_{-0.35}$ \\
2 & $15.50^{+0.50}_{-0.40}$ & $15.80^{+0.60}_{-0.40}$ & $14.65^{+0.30}_{-0.35}$ & $14.30^{+0.20}_{-0.25}$ & $14.35^{+0.15}_{-0.15}$ \\
3 & $15.20^{+0.45}_{-0.40}$ & $15.40^{+1.10}_{-0.50}$ & $14.70^{+0.35}_{-0.35}$ & $<15.00$ & $14.25^{+0.15}_{-0.30}$ \\
4 & $15.00^{+0.15}_{-0.40}$ & $14.80^{+0.30}_{-0.30}$ & $<13.90$                & -- & $15.40^{+0.30}_{-0.35}$  \\
5 & $14.90^{+0.60}_{-0.50}$ & $<14.60$                & $<14.60$                & -- & $<14.45$  \\
\noalign{\smallskip}\hline\noalign{\smallskip}
H$_2$ (total) & $17.26^{+0.16}_{-0.19}$ & $17.10^{+0.44}_{-0.53}$ & $15.34^{+0.15}_{-0.10}$ & $15.38^{+0.40}_{-0.10}$ & $16.12^{+0.24}_{-0.20}$ \\
\noalign{\smallskip}\hline\noalign{\smallskip}
$T_{01}$ [K] & $55^{+43}_{-16}$ & $42^{+249}_{-16}$ & $82^{+127}_{-29}$ & $>37$ & $55^{+68}_{-17}$ \\
$T_{02}$ [K] & $96^{+47}_{-18}$ & $117^{+438}_{-38}$ & $>185$ & $200^{+430}_{-100}$ & $101^{+37}_{-19}$ \\
$T_{24}$ [K] & $670^{+1800}_{-370}$ & $400^{+500}_{-170}$ & $<780$ & -- & -- \\
$T_{35}$ [K] & $>450$ & $\la650$ & $\la2200$ & -- & $\approx4400$ \\
\noalign{\smallskip}\hline
\end{tabular}
\end{flushleft}
\end{table*}

\begin{table*}
\caption[]{Halo gas H$_2$ column densities $\log N$ [cm$^{-2}$] and 
      H$_2$ excitation temperatures towards the LMC stars. 
      For the three rightmost sight lines the first value gives the column 
      density for a Doppler parameter $b=1$~km\,s$^{-1}$, the second assumes         the equivalent width  at the Doppler limit of the curve of growth, which       is quite appropriate  for $b$-values larger than about $5$~km\,s$^{-1}$}
\label{H2CD_halo}
\begin{flushleft}
\setlength{\tabcolsep}{1.8mm}
\begin{tabular}{llllllllll}
\hline\noalign{\smallskip}
J & \multicolumn{2}{l}{\object{\object{HD~269546}}} & \multicolumn{2}{l}{LH 10:3120} & \multicolumn{2}{l}{\object{Sk~-67\,166}} & \multicolumn{2}{l}{\object{HD~36402}} \\
 & IVC & HVC              & IVC & HVC       & IVC & HVC       & IVC & HVC \\ 
\noalign{\smallskip}\hline\noalign{\smallskip}
$v_{\mathrm{hel}}$ [km\,s$^{-1}$]& $+60$ & $+120$ & $+60$ & $+120$ & $+60$ & $+120$ & $+60$ & $+120$ \\
\noalign{\smallskip}\hline\noalign{\smallskip}
0 & $15.30^{+0.50}_{-0.40}$ & $<14.20$                & $<15.6/14.0$ & $<18.4/14.5$ & $<15.3/14.1$ & $<14.6/13.8$ & $<14.4/13.7$ & $<13.6/13.5$ \\
1 & $15.20^{+0.40}_{-0.35}$ & $14.60^{+0.20}_{-0.20}$ & $<18.3/14.7$ & $<18.2/14.6$ & $<16.0/14.1$ & $<17.6/14.3$ & $<14.5/13.9$ & $<16.6/14.0$ \\
2 & $<14.40$                & $14.80^{+0.20}_{-0.20}$ & $<17.3/14.4$ & $<17.5/14.3$ & $<14.4/13.8$ & $<13.9/13.6$ & $<13.9/13.6$ & $<13.9/13.6$ \\
3 & $14.80^{+0.05}_{-0.70}$ & $14.80^{+0.20}_{-0.20}$ & $<16.0/14.3$ & $<16.1/14.3$ & $<14.8/14.0$ & $<16.4/14.2$ & $<15.7/14.3$ & $<16.3/14.2$ \\
4 & --                      & $14.70^{+0.40}_{-0.40}$ & --           & --       & --       & --       & -- & -- \\
5 & --                      & $<14.70$                & --           & --       & --       & --       & -- & -- \\
6 & --                      & $<14.90$                & --           & --       & --       & --       & -- & -- \\
\noalign{\smallskip}\hline\noalign{\smallskip}
H$_2$ (total) & $15.65^{+0.32}_{-0.19}$ & $15.56^{+0.10}_{-0.06}$ & $<18.3/15.0$ & $<18.6/15.0$ & $<16.1/14.6$ & $<17.6/14.7$ & $<15.8/14.6$ & $<16.8/14.5$ \\
\noalign{\smallskip}\hline\noalign{\smallskip}
$T_{\mathrm{ex}}$ [K] & $<150$ & --                    & -- & -- & -- & -- & -- & -- \\
$T_{\mathrm{UV}}$ [K] & --    & $>1000$               & -- & -- & -- & -- & -- & -- \\
\noalign{\smallskip}\hline
\end{tabular}
\end{flushleft}
\end{table*}

\begin{table*}
\caption{Column densities of metals and \ion{H}{i} as $\log N$ [cm$^{-2}$] towards the four program stars.  $v_{\rm hel}$ [km\,s$^{-1}$] is the heliocentric radial velocity of the absorption components from the {\sc orfeus} data, $b$ [km\,s$^{-1}$] is the Doppler parameter of the curves of growth}     
\label{colden} 
\setlength{\tabcolsep}{0.8mm}
\begin{tabular}{lllllllllllll}
\hline\noalign{\smallskip}
  &\multicolumn{3}{l}{\object{\object{HD~269546}}}&\multicolumn{3}{l}{LH 10\,3120}&\multicolumn{3}{l}{Sk -67\,166}&\multicolumn{3}{l}{\object{HD~36402}}  \\ \noalign{\smallskip}\cline{2-4}\cline{5-7}\cline{8-10}\cline{11-13}\noalign{\smallskip}
 & disk & IVC & HVC & disk & IVC & HVC & disk & IVC & HVC & disk & IVC & HVC\\ 
\noalign{\smallskip}\hline\noalign{\smallskip}
$v_{\rm hel}$ & $-9^{+3}_{-3}$ & $+56^{+7}_{-7}$ & $+118^{+4}_{-4}$ & $-16^{+2}_{-2}$ & $+51^{+5}_{-5}$ & $+122^{+10}_{-10}$ & $-5^{+3}_{-3}$ & $+61^{+3}_{-3}$ & $+125^{+5}_{-5}$ & $+15^{+4}_{-4}$ & $+70^{+4}_{-4}$ & $+131^{+3}_{-3}$ \\
\noalign{\smallskip}\hline\noalign{\smallskip}
 $b$ & $22^{+3}_{-3}$ & $7^{+3}_{-3}$ & $11^{+3}_{-3}$ & - & - & - & $11$-$19$ & $5^{+2}_{-2}$ & $22^{+3}_{-3}$ & $15^{+5}_{-5}$ & $9^{+2}_{-2}$ & $18^{+7}_{-6}$ \\
\noalign{\smallskip}\hline\noalign{\smallskip}
\ion{O}{i}  & $16.35^{+0.3}_{-0.3}$ & $14.3^{+0.5}_{-0.9}$ & $15.2^{+0.3}_{-0.2}$ & $>$15.8 & -  & $>$15.0 & $16.2^{+0.3}_{-0.5}$ & $15.0^{+1.2}_{-0.7}$ & $14.85^{+0.2}_{-0.2}$ & $16.2^{+0.3}_{-0.3}$ & $15.25^{+0.3}_{-0.3}$ & $15.2^{+0.3}_{-0.2}$ \\
\ion{Si}{ii} & $15.4^{+0.15}_{-0.15}$ & $14.6^{+0.5}_{-0.5}$ & $15.1^{+0.3}_{-0.2}$ & $>$13.9 & $>$13.5 & $>$13.6 & $15.35^{+0.15}_{-0.10}$ & $14.0^{+0.5}_{-0.7}$ & $14.3^{+0.3}_{-0.2}$ & $15.35^{+0.15}_{-0.15}$ & $14.0^{+0.5}_{-0.5}$ &$14.6^{+0.3}_{-0.3}$ \\  
\ion{Fe}{ii} &$15.0^{+0.2}_{-0.2}$ & $14.5^{+0.3}_{-0.3}$ & $14.4^{+0.3}_{-0.3}$ & $>$15.1 & $>$13.5 & $>$14.2 & $14.85^{+0.2}_{-0.5}$ & $14.4^{+0.4}_{-0.6}$ & $14.1^{+0.4}_{-0.3}$ & $14.5^{+0.2}_{-0.2}$ & $14.2^{+0.5}_{-0.4}$ & $14.2^{+0.3}_{-0.4}$  \\
\ion{S}{ii}  & $15.2^{+0.2}_{-0.2}$ & $14.2^{+0.4}_{-0.2}$ & $14.65^{+0.3}_{-0.3}$ & $>$15.0 & $>$15.0 & $>$15.1 & $15.3^{+0.2}_{-0.2}$ & $14.8^{+1.5}_{-1.5}$ & $14.85^{+0.2}_{-0.2}$ & $15.1^{+0.2}_{-0.2}$ & $14.3^{+0.4}_{-0.2}$ & $14.7^{+0.4}_{-0.2}$ \\
\ion{Mg}{i} & $12.6^{+0.3}_{-0.2}$ & $12.6^{+1.4}_{-0.3}$ & $12.8^{+0.7}_{-0.4}$ & - & - & - & $12.4^{+0.2}_{-0.2}$ & $11.9^{+0.6}_{-0.4}$ & $12.25^{+0.2}_{-0.2}$ & $12.6^{+0.5}_{-0.3}$ & $12.3^{+0.5}_{-0.3}$ & $12.3^{+0.4}_{-0.3}$ \\ 
\ion{Mg}{ii} & $15.45^{+0.7}_{-0.6}$ & $12.85^{+0.8}_{-0.9}$ & - & - & - & - & $14.5^{+1.0}_{-0.6}$ & $14.3^{+1.2}_{-0.8}$ & $13.8^{+1.2}_{-0.5}$ & $15.6^{+0.9}_{-1.2}$ & $13.5^{+0.9}_{-0.5}$ & $14.2^{+0.8}_{-0.5}$ \\
\ion{H}{i} & $20.54^{+0.07}_{-0.08}$$_r$ & $18.2_w$ & $19.05_w$ & \multicolumn{2}{l}{$20.7^{+0.8}_{-0.8}$$_{r}$} & & $20.54^{+0.08}_{-0.09}$$_r$ & $<18.2_p$ & $<18.0_p$  & $20.73_s$ & $18.64_w$ & $18.64_w$ \\ 
\noalign{\smallskip}\hline\noalign{\smallskip}  
\end{tabular} 
\noindent

$w$ from Wierig \& de Boer (\cite{wierig}) based on Parkes 15$^{\prime}$ beam 21 cm data by McGee et\,al.\, (1983) and McGee \& Newton (1986) \\
$r$ from Richter (\cite{richter2000}) \\
$p$ from a Parkes 15$^{\prime}$ beam 21 cm emission profile (centered $\approx14^{\prime}$ off the position of \object{Sk~-67\,166}) by McGee \& Newton (1986) \\
$s$ from Savage \& de Boer (1981)\\
Note that the \ion{O}{i} column densities in disk gas have to be treated as lower limits (see Sec. 4.1)
\end{table*}

\subsection{Disk gas}

\begin{figure}
\resizebox{\hsize}{!}{\includegraphics{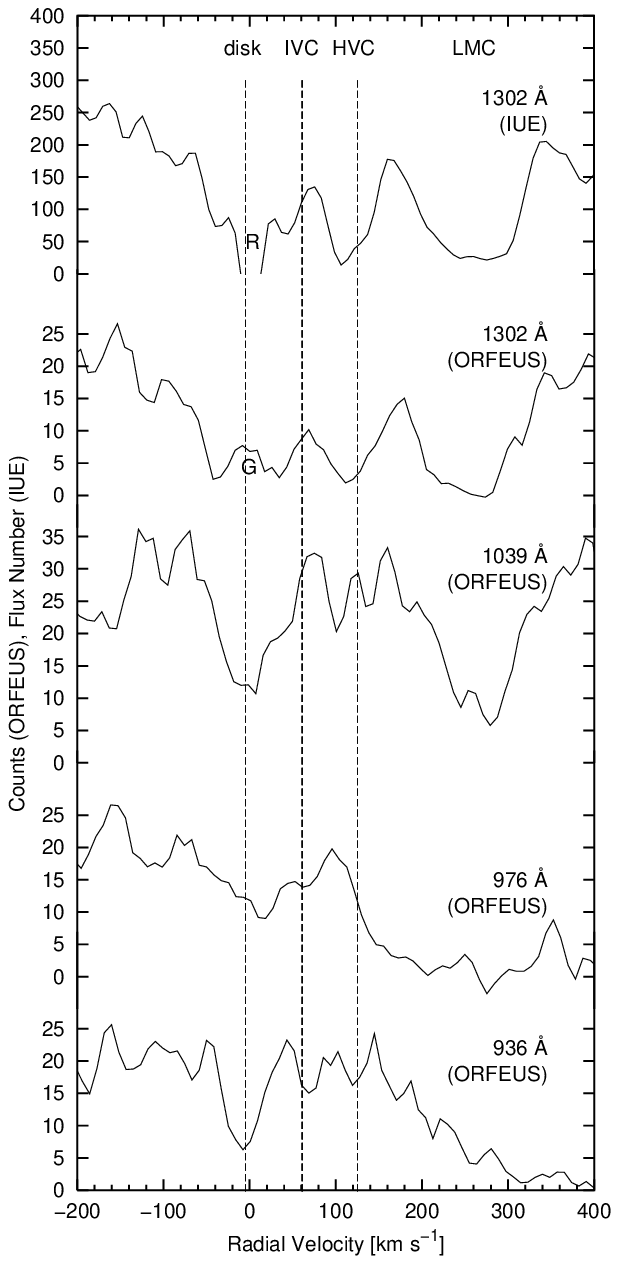}}
\hfill
\caption[]{Profiles of some \ion{O}{i} lines in the spectrum of Sk~$-67\,166$. The radial velocity  is heliocentric and not corrected for possible pointing errors. Note the reseaux (R) at the centre of the $1302$~{\AA} line in the {\sc iue}~spectrum and the geocoronal emission (G) at the same place in the corresponding {\sc orfeus}~spectrum. Significant contamination of other \ion{O}{i} lines by airglow is not as obvious, but cannot be excluded  }
\label{oxy}
\end{figure}

\begin{figure*}
\resizebox{\hsize}{!}{\includegraphics{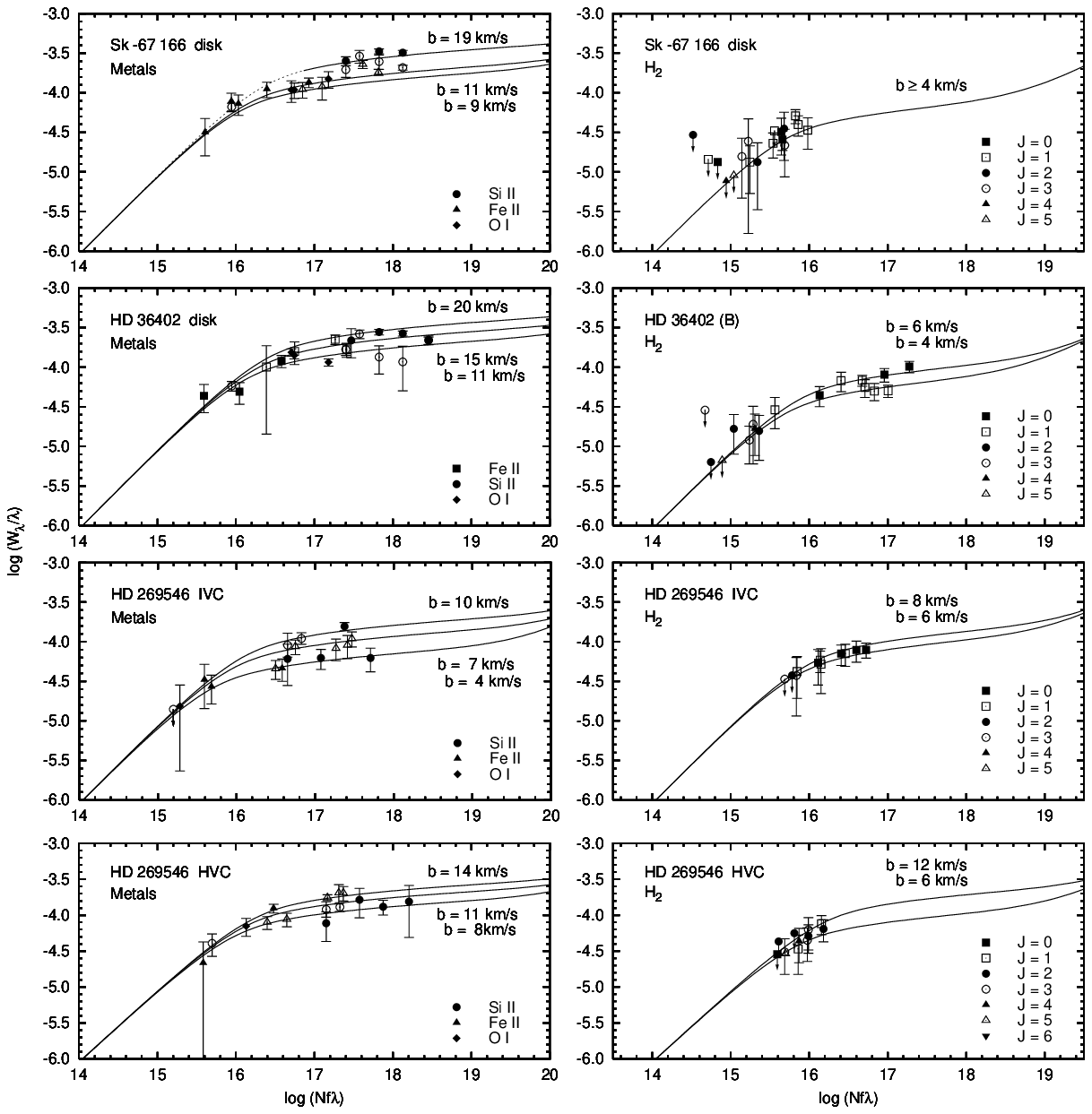}}
\hfill
\caption[]{Examples for the curves of growth for absorption by metals (left) and H$_2$ (right) are shown. For the metals, filled symbols represent {\sc ORFEUS} data, open symbols those from the {\sc IUE}. The drawn curves represent single cloud absorption with indicated $b$-value }
\label{cogs}
\end{figure*}

 Gas near 0 km\,s$^{-1}$ is gas in the local Milky Way disk, 
whose thickness is normally indicated by a scale height of $\simeq 100$ pc. 
 However, gas is known to exist slightly further out, 
at distances up to 1 kpc (Lockman et al. \cite{lockman}). 
 The four {\sc orfeus} lines of sight span about 3$^{\circ}$ on the sky which, 
at a distance of less than 1 kpc, 
represents a lateral extent of less than 60 pc. 

 The column densities of metal species do not show significant variations between the different lines of sight, except perhaps for \ion{Fe}{ii} (see Table \ref{colden}).
 For oxygen we have to note that the absorption lines in the {\sc orfeus} spectra are contaminated by geocoronal emission. 
 The absorption profile is partly filled in by the emission so that the measured equivalent widths of the disk component are lower limits, while the IVC and HVC lines are not affected. 
 This explains why the \ion{O}{i} disk depletions are given as lower limits in Table~\ref{depletion}.  

 The absorption by the disk gas probably consists of at least two unresolved components.
 This is supported by the high Doppler-parameters ($b$ between $15$ and $22$~km\,s$^{-1}$) of the metal curves of growth. 
 A clear deviation from the single cloud curve is visible in the rise in the level of the flat part of the curve of growth of the Sk~-67\,166 disk component.  
 The H$_2$ measurements support the suspicion of the presence of more than one component, since for this sight-line the radial velocity of H$_2$ lies asymmetric within the velocity width of the metal lines.
 Towards \object{HD~36402} two H$_2$ absorption components (A and B) on either side of the mean metal radial velocity are detected. 
 The spectra of \object{LH~10:3120} are too noisy to recognize the double nature of the absorption.
 Optical \ion{Ca}{ii} spectra at higher resolution (as shown e.g. in Wayte \cite{wayte}) towards other LMC stars confirm the existence of velocity structure in disk gas in this direction.

 The metals are moderately depleted in the disk component, e.g. $\log$(S/H)$-\log$(S/H)$_{\sun} = -0.5$ to $-0.8$ and $\log$(Fe/H)$-\log$(Fe/H)$_{\sun} = -1.0$ to $-1.7$.  
 This points to the existence of dust on these lines of sight, but not in a large quantity.

 The H$_2$ column densities for the local gas (see Table~\ref{H2CD_local}) scatter more than the \ion{H}{i} and metal column densities.
However, this is not surprising. 
H$_2$ is expected to exist mainly in the cores of interstellar clouds, surrounded by the more extended atomic gas, and our lines of sight surely sample different parts of the molecular clouds.

 The H$_2$  Boltzmann excitation temperatures were determined for the lower levels $J=0$ and $1$ as well as for $0$ and $2$.
 While the $J=0,1$ comparison is directly linked to the ortho-H$_2$ to para-H$_2$ ratio (OPR), which might not be fully thermalized, the $J=0,2$ comparison refers to the two lowermost para-states and thus gives the real kinetic temperature (at least as long as collisional excitation predominates in both states).

To investigate the upper UV pumping excited levels, we use the equivalent temperatures for the para-states $2$ and $4$ and for the ortho-states $3$ and $5$, respectively.
The temperatures are listed in Table \ref{H2CD_local}, for some of the values only limits can be given.

 For all lower states the Boltzmann excitation temperatures are found within a range of about $50$ to $100$~K, as also found by Savage et al. (\cite{savage77}) for general galactic gas from {\sc Copernicus} data.
 For HD\,269546, LH\,10:3120 and for component B of HD\,36402 $T_{01}$ and $T_{02}$ are consistent within the range of errors. 
 The gas on these lines of sight thus should be fully thermalised.
 For the sight lines towards Sk\,-67\,166 and HD\,36402 (A) $T_{02}$ is significantly higher than $T_{01}$. 
 In  this dilute, well illuminated gas the $J=2$ states presumably are already excited by UV pumping, so that they are overpopulated compared to what would be expected from just collisional excitation.
 Here we also find high excitation temperatures of the upper states, again due to the low density preventing efficient self shielding.
 Thus a clear distinction between collisionally excited states and states excited by UV pumping is not possible.

 The column densities on all sight lines are quite similar for the upper, UV pumping excited $J$ levels. 
Only the lower, collisionally excited levels vary significantly in column density.
This allows the conclusion that we look in the direction of the LMC through a layer of quite homogeneously distributed UV pumping excited molecular gas, with regions of differing column densities in the cooler lower states.

 The H$_2$ absorption component B towards \object{HD~36402} at $+35$~km\,s$^{-1}$ 
seems to be compact enough to contain a shielded core at a Boltzmann temperature of about $T_{\rm ex}<100$~K.
 Similar  conditions are found on the two other high column density sight lines, towards HD\,269546 and LH\,10:3120. 
 The upper excitation levels indicate a high equivalent temperature of $T_{\rm UV}>1000$~K due to UV pumping,  occuring probably in the outer cloud regions.

 The atomic absorption found near $v_{\rm hel}\simeq+15$~km\,s$^{-1}$ in the \object{HD~36402} spectrum is rather wide (from $-15$ to $+45$~km\,s$^{-1}$ ) and is, as noted before, probably a blend of the two velocity components seen in H$_2$ at $v_{\rm hel}\simeq-5$~km\,s$^{-1}$ and $v_{\rm hel}\simeq+35$~km\,s$^{-1}$.
 An exception is the 2852~{\AA} line of \ion{Mg}{i}, 
a tracer of cold gas.
 This line is narrow ($FWHM\approx30$~km\,s$^{-1}$) and shows no sub-structure. 
 It belongs to the $+35$~km\,s$^{-1}$ component, as follows immediately from a comparison of the \ion{Mg}{i} absorption profile with that of \ion{Mg}{ii} $\lambda$2802. 
 From this we conclude that component B contains gas which is cooler and denser than that in component A.

%
\subsection{The IVC gas}

 We have detected H$_2$ absorption in the IVC gas ($v_{\rm hel}\approx+56$~km\,s$^{-1}$) towards \object{HD~269546}. 
 On the other lines of sight we have significant upper limits.  
 The distribution of the column densities over the various rotationally excited levels of H$_2$ (see Table\,\ref{H2CD_halo}) can be used to calculate the kinetic temperature of the gas. 
 Using the two para states $J=0$ and $2$ we find an upper limit of $T_{02} < 150$ K. 
 Level $J=3$ has been detected in absorption, too, but the uncertainty in the column density is large. 
Still,  its detection suggests that the UV photon field is quite important in the IV gas. 
 In fact the full continuum flux from the Milky Way disk is available for photo-excitation, unattenuated by H$_2$ line absorption (see Appendix).

 Intermediate velocity gas is found in many high latitude directions 
in the Milky Way (see Danly \cite{danly}). 
 As noted in the introduction, the present detection of H$_2$ absorption in IVC gas is the third case of H$_2$ proven to be existent in such gas.

 On our four lines of sight IVC absorptions in the metals are seen at $v_{\rm hel} = 50$ to $70$ km\,s$^{-1}$ (see Table\,\ref{H2CD_halo}).
 These absorptions are weaker than those by disk and HVC gas.
 They are often poorly resolved from these other velocity components, so the uncertainties in the IVC equivalent widths are large.
 Metal column densities appear to be large with respect to \ion{H}{i} (see Table\,\ref{colden}), except for oxygen.  
 This abundance pattern is also seen in the HVC gas (see next section).

 The relative absence of O (see also Table\,\ref{depletion}) can be interpreted as due to partial photoionization of the gas, affecting predominantly H and O, and less the other elements, because the ionization potential of \ion{H}{i} and \ion{O}{i} is $13.6$~eV, lower than that of \ion{Si}{II} ($16.3$~eV), \ion{S}{II} ($23.33$~eV) or \ion{Fe}{ii} ($16.2$~eV).
 \ion{O}{ii} does not have absorption lines in the UV.
 Therefore we estimate the $N$(\ion{O}{ii}) to $N$(\ion{O}{i}) ratio by comparing the \ion{O}{i} and \ion{S}{ii} column densities.
 Here we use the fact that the oxygen and sulfur abundance ratio is only weakly related to the metallicity of the gas (see for example Samland \cite{saml}) as well as that the oxygen and sulfur abundances are barely affected by deposition onto dust grains.
 In the neutral interstellar medium the common value of $\log$[$N$(O)/$N$(S)] is $\approx1.2\pm{0.2}$ (for the \ion{O}{i} abundance in the neutral ISM see Meyer et al. \cite{meyer}).
 In contrast, towards \object{HD~269546}, we find $\log [N$(\ion{O}{i})/$N$(\ion{S}{ii})$]=0.1^{+0.5}_{-1.0}$.
 Thus the actual oxygen column density is presumably higher by {\bf $1.1$} dex than that of \ion{O}{i}. 
 For temperatures above $\approx1000$~K the ionization level of oxygen is coupled to that of hydrogen due to charge exchange reactions (see e.g. Field \& Steigman \cite{fiestei}). 
 Because of  $N$(\ion{H}{ii})/$N$(\ion{H}{i})$\approx\frac{9}{8}N$(\ion{O}{ii})/$N$(\ion{O}{i}), the hydrogen column density has to be increased by the same amount as the oxygen column density, which would mean an ionization level of more than $90~\%$ if our assumptions are correct. 
 The line of sight to \object{HD~36402} is different, it has rather normal abundances and $N$(\ion{O}{i})/$N$(\ion{S}{ii}) indicates no ionization.

 On most sight lines (except towards HD\,269546) noise and blending only allow to derive upper limits for the H$_2$ content in the IVC gas (see Table \ref{H2CD_halo}). 
 To obtain these we fitted the upper limits for the equivalent widths to curves of growth with a $b$-value of $1$~km\,s$^{-1}$ and to the Doppler limit of the curve of growth.
 The lower $b$-value should give a distinct limit in the case of very cold gas cores.
 The Doppler limit largely applies for all $b$-values $\geq5$~km\,s$^{-1}$.
 Experience shows that in gas this dilute $b$-values are normally significantly larger than $1$~km\,s$^{-1}$, similar to metal $b$-values.

 The metal and H$_2$ absorptions together indicate that
the fraction of H$_2$ of all H is rather small, near $10^{-3.5}$, when including the correction for fractional ionization of H of 90 \% as estimated above. 
 These lines of sight sample IV gas which has dilute and ionized portions as well as cooler and denser portions. 

 Wolfire et al. (\cite{wolfire}) investigated the pressure range for which stable two-phase gas can exist in the galactic halo.
 For cloud heights between 1~kpc and 3~kpc their models give $P$/k values of  $10^{3.2}$~K\,cm$^{-3}$ to $10^{3.7}$~K\,cm$^{-3}$.
 Under the assumption of a pressure of about $10^{3.5}$~K\,cm$^{-3}$ and with the upper limit for the kinetic temperature of H$_2$, it is possible  to estimate as lower limit $n_{\rm H} \ga 20$~cm$^{-3}$ for the particle density in the molecular core of the IVC.  
 The \ion{H}{i} column density divided by this number density yields a size of $\approx0.025$~pc.
 This is an upper limit for the cold core of the cloud. 
It is at the same time a lower limit for the total cloud since the outer regions are certainly warmer than 150~K.

\begin{table}
\caption[]{$\log$[X/\ion{H}{i}]-$\log$[X/H]$_{\sun}$ in disk, IVC, and HVC gas. Solar abundances are taken from Savage \& Sembach (\cite{savsem}). The uncertainties are at least as large as those of the metal column densities. The overabundances of some elements relative to \ion{H}{i} and \ion{O}{i} indicate a significant degree of ionization in the HVC and the IVC. The lower limits for oxygen in disk gas reflect the uncertainties due to geocoronal \ion{O}{i} emission at the velocity of the disk component}
\label{depletion}
\begin{tabular}{llccccc}
\hline\noalign{\smallskip}
 &  & O & S & Si & Fe & Mg \\
\noalign{\smallskip}\hline\noalign{\smallskip}
 & \object{\object{HD~269546}} & $\geq$-1.2 & -0.5 & -0.6 & -1.0 & -0.7 \\
disk & Sk -67\,166 & $\geq$-1.2 & -0.5 & -0.7 & -1.3 & -1.6 \\
 & \object{HD~36402} & $\geq$-1.4 & -0.8 & -0.9 & -1.7 & -0.7 \\
\noalign{\smallskip}\hline\noalign{\smallskip}
 & \object{\object{HD~269546}} & -0.8 & +0.8 & +1.0 & +0.8 & +0.8 \\
 \raisebox{1.5ex}[-1.5ex]{IVC} & \object{HD~36402} & -0.3 & +0.5 & -0.1 & +0.1 & -0.7 \\
\noalign{\smallskip}\hline\noalign{\smallskip}
 & \object{\object{HD~269546}} & -0.7 & +0.4 & +0.5 & -0.2 & - \\
 \raisebox{1.5ex}[-1.5ex]{HVC} & \object{HD~36402} & -0.3 & +0.9 & +0.5 & +0.1 & -0.0 \\
\noalign{\smallskip}\hline
\end{tabular}
\end{table}

\subsection{The halo HVC gas}

\begin{figure}
\resizebox{\hsize}{!}{\includegraphics{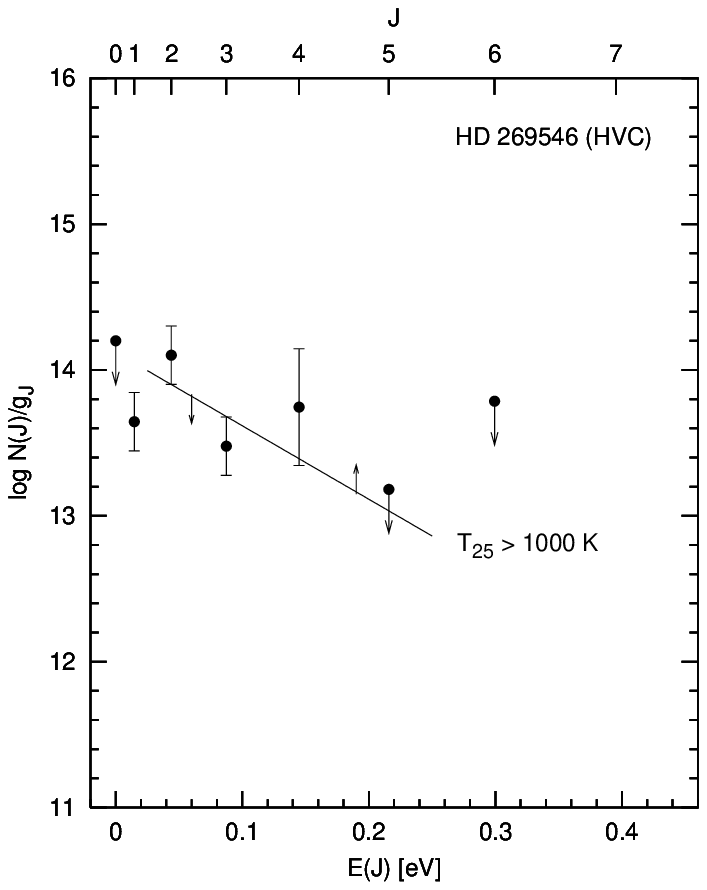}}
\hfill
\caption[]{Excitation plot for the H$_2$ HVC component towards HD\,269546. 
The alternation in column densities is due to the ortho-to-para ratio of 
$\simeq1$. 
The excitation temperature due to UV pumping is derived to be $>1000$~K}
\label{T_HVC}
\end{figure}

 The HVC gas on the LMC line of sight was known (as was the IVC gas) since the first {\sc iue} observations (Savage \& de~Boer \cite{savdB}). 
 Absorption by H$_2$ was detected in {\sc orfeus} spectra  towards \object{HD~269546} 
and essentials have been presented by Richter et\,al.\, (\cite{richter99a}). 
 Here we will discuss also aspects which could not be covered in that paper. 

 The distance of the HVC is still unknown. 
 However, HVCs have probably typical distances of $z\ga3$ kpc 
(Kaelble et\,al.\, \cite{kaelble}; van Woerden et\,al.\, \cite{vW}; 
see also review by Wakker \& van Woerden \cite{wakker}). 
 Taking that distance in $z$ implies that the {\sc orfeus} lines of sight 
intersect the +120~km\,s$^{-1}$ HVC over a projected width of $\ga350$ pc.
 The three sight-lines for which metal column densities can be given correspond to a width of $\ga 130$~pc in the assumed distance.
 Though the Doppler parameters have values between $11$ and $22$~km\,s$^{-1}$, the column densities are similar towards the different targets.  
   
 On all lines of sight \ion{O}{i} is depleted relative to the other metals.
 Applying the ionization correction described in the previous section, 
we find $\log N$(\ion{H}{ii})/$N$(\ion{H}{i})$\approx \log N$(\ion{O}{ii})/$N$(\ion{O}{i}) as $0.7$, $1.2$, and $0.7$ dex towards \object{HD~269546}, Sk~-67\,166, and \object{HD~36402} respectively.  
 This means that in the HV gas roughly only $5$ - $20$\,\% of the hydrogen is neutral.

 The column densities of metals are given in Table\,\ref{colden}, the apparent depletions (without ionization correction) in Table\,\ref{depletion}. 
 Taking S as the reference for the abundances we find that Si and Fe are depleted by factors ranging from 2 to 4. 
 Based on the Parkes \ion{H}{i} column densities and the ionization correction, the abundances of S, Si, and Fe turn out to be similar to those in the warm local ISM.
 Of course there are considerable uncertainties in the ionization correction as well as in the \ion{H}{i} column densities (because of the Parkes beam size). 
 Given the metal content neither a galactic nor an LMC origin of this HVC can be excluded.

\begin{figure}
\resizebox{\hsize}{!}{\includegraphics{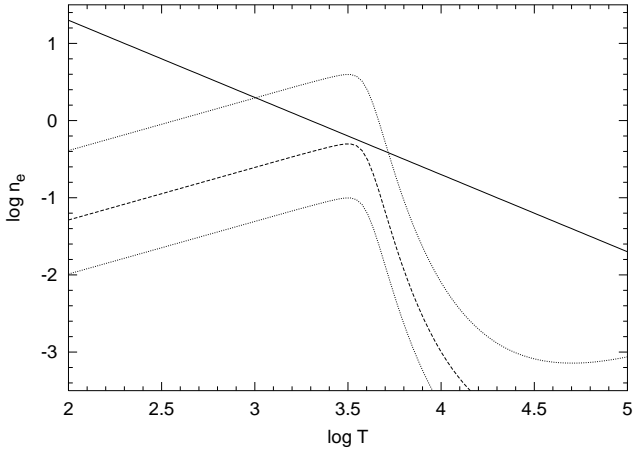}}
\hfill
\caption[]{For the HVC on the line of sight to \object{HD~36402} we show $n_e(T)=\frac{\Gamma}{\alpha(T)}\cdot\frac{N({\rm \ion{Mg}{i}})}{N({\rm \ion{Mg}{ii}})}$ (dashed curve, with dotted curves representing the errorbars of $\frac{N({\rm \ion{Mg}{i}})}{N({\rm \ion{Mg}{ii}})}$).
 The straight line gives $n_e=\frac{P_e}{k}\cdot\frac{1}{T}$  for the pressure $n_eT=10^{3.3}$cm$^{-3}$K, as indicated by the models of Wolfire et al. (\cite{wolfire}) for reasonable $z$ heights. The intersection allows an estimate of density and temperature in the HVC }
\label{hvcion}
\end{figure}

 With the \ion{Mg}{i} to \ion{Mg}{ii} ratio it is possible to estimate temperature, density, and extent of the HVC.
 We assume a $z$ height between $3$~kpc and $10$~kpc and a pressure $\log(P/{\rm k})\approx3.3$ (a value from the models of Wolfire et al. \cite{wolfire}).
 If the gas is in ionization equilibrium and if charge exchange reactions are neglegible,  $N$(\ion{Mg}{ii})/$N$(\ion{Mg}{i}) is determined by $\frac{\Gamma}{\alpha n_e}$, with the photoionization rate $\Gamma$, the temperature dependent recombination coefficient $\alpha$, and the electron density $n_e$.
 Fits for the different contributions to $\alpha$ can be found in P\'equignot \& Aldrovandi (\cite{peqaldro}), Shull \& Van Steenberg (\cite{shull}) and Nussbaumer \& Storey (\cite{nuss}).
 The flux of ionizing photons and the resulting photoionization rates at the location of the HVC are unknown.
 We use the photoionization rate for \ion{Mg}{i} in the galactic disk given in de Boer et al. (\cite{dBkopp}), scaled down by a factor of $\approx 3$ to the halo radiation field as estimated by Bregman \& Harrington (\cite{bregman}).  
 In Fig. \ref{hvcion} we plot $n_e(T)=\frac{\Gamma}{\alpha(T)}\cdot\frac{N({\rm \ion{Mg}{i}})}{N({\rm \ion{Mg}{ii}})}$ and $n_e=\frac{P_e}{k}\cdot\frac{1}{T}$ for the HVC gas on the \object{HD~36402} line of sight.
 The intersection of these curves provides information about temperature and electron density in the cloud.
 It indicates $0.08 < n_e < 3$~cm$^{-3}$ at temperatures $1000 < T < 6000$ K.
 Because of the high degree of ionization the dominant source of electrons is hydrogen and $n_e$ equals $n$(H$^+$) $\approx5{\cdot}n$(\ion{H}{i}).
 With the value for the column density $N$(\ion{H}{i})= $4.4 \cdot 10^{18}$ cm$^{-2}$, we now can estimate the extent of the cloud along the line of sight to be of the order of $1-100$~pc.

 In the HVC gas H$_2$ is only found towards \object{HD~269546} (Richter et al. \cite{richter99a}). 
 Here the $b$-value for H$_2$ is $9$~km\,s$^{-1}$, only $2$~km\,s$^{-1}$ lower than that for the metals.
 As with the IVC, we can assume that the full disk radiation field operates on HVC gas, too (see App.).
 The excitation temperature ($\geq 1000$~K, see Fig. \ref{T_HVC}) and the ortho-to-para ratio ($\approx 1$) of H$_2$ point to deviations from thermodynamic equilibrium conditions in this part of the HVC.
 This can be explained by quite different mechanisms, such as H$_2$ formation or a former cloud core, which is now exposed to and photodissociated by UV flux from the galactic disk.   
 Spitzer \& Zweibel (\cite{spitzweib}) calculated the steady state H$_2$ photo excitation for different radiation fields, densities, and temperatures.
 It turns out that none of their models yields excitation temperatures as high as measured in the HVC, except for the case of recent H$_2$ formation.
 Though it is conceivable that in the galactic fountain model H$_2$ is being formed in the cooling gas falling back to the galactic plane, the present data do not allow a full assessment of the physical conditions in this cloud.

 As for the IVC, also for the HVC only upper limits are available for H$_2$ towards the remaining targets (Table~\ref {H2CD_halo}), again derived assuming $b$-values of $1$~km\,s$^{-1}$ and $b>5$~km\,s$^{-1}$.
 In most cases the deviation between the upper limits of para-H$_2$ states and their corresponding ortho-states is much larger than one would expect for typical H$_2$ excitation. 
 Thus, if ortho- and para-states are both taken into account, the upper limits for many of the excitational states are smaller than stated in Table \ref{H2CD_halo}.

\section{Overall conclusions}
 It has been suggested that HVCs are building blocks of the Local Group containing low metallicity gas (e.g. Blitz et al. \cite{blitz}).
 The IVC and the HVC in front of the LMC do not fit in this model.
 Metal abundances and the existence of H$_2$ support the galactic fountain model for this IVC/HVC complex.
 Though the underabundance of metals in the HVC seems superficially consistent with a Magellanic Cloud origin of that gas, the presence of H$_2$ points to the existence of dust and thus intrinsically higher metallicity.

\begin{appendix}
\section*{Appendix: Radiation field in the halo and H$_2$ photo excitation}

 In order to interpret the level of photo excitation 
of H$_2$ in the halo clouds one has to know the radiation field. 
 In the lower halo the radiation density is essentially the same 
as that in the Milky Way (MW) disk. 
 Very far out the radiation field is diluted because the gas will not see 
so large a solid angle of Milky Way radiation. 

 What matters for H$_2$ and its photo excitation is 
the amount of photons at the wavelength of the H$_2$ absorption lines. 
 Nominally, these photons have been absorbed in the molecular gas 
of the MW disk. 
 A halo cloud will, however, 
see radiation from different parts of the Milky Way disk 
having different radial velocities due to the galactic rotation. 
 Moreover, halo clouds will normally not have zero velocity 
with respect to the MW disk. 

 We have performed a simple integration over the galactic disk to 
determine the distribution of the velocity shifts. 
 A map of the MW with model radial velocities was divided 
in 20$^{\circ}$ sectors overlain with circles with steps of 1~kpc in radius. 
 For each of these curved boxes the average LSR radial velocity 
was estimated by eye. 
 All these values, 
weighted with the surface area of the box and with the distance, 
lead to a histogram of distance related velocity shifts. 
 From the histogram it is immediately clear, 
that $>75$\% of the radiation from the disk 
arrives in the halo with a shift of less than 30 km\,s$^{-1}$. 
 Halo clouds with a velocity near 0 km\,s$^{-1}$ will thus recieve $<25$ \% of 
the continuum flux level for photo excitation. 
 However, since the great majority of clouds detected in the halo, 
IVCs and HVCs alike, 
have a radial velocity differing by more than 30 km\,s$^{-1}$ from the LSR, 
the H$_2$ in these halo clouds sees almost the full continuum radiation 
at the wavelengths of the H$_2$ absorption lines, 
i.e., unattenuated by the H$_2$ line disk absorption.
 Almost the full continuum flux is available for photo excitation. 

\end{appendix}

%

%
\acknowledgements
We thank the {\sc orfeus} team for making and operating {\sc orfeus} 
and providing us with the data.
HB is supported by the GK {\it The Magellanic Clouds and other dwarf galaxies}, PR was supported by grant 50 QV 9701-3 from the DARA (now DLR), 
OM by grant Bo 779/24 from the Deutsche Forschungsgemeinschaft (DFG). 
Part of the data has been collected under the 
DARA {\sc orfeus} guest observer programme. 
We also made use of the public archive of IUE spectra.

\end{document}